\documentclass{PoS}
\usepackage{amsmath}
\usepackage{booktabs}
\usepackage{caption}
\usepackage{bm}  
\usepackage{cite}
\usepackage{soul}
\usepackage{float}

\usepackage{graphics}
\usepackage{subfig}
\usepackage{slashed}

\newcommand{\BL}{L+R}

\renewcommand{\i}{\mathrm i}

\newcommand{\U}[1]{\mathrm{U}(1)_{\mathrm{#1}}}			
\newcommand{\SU}[2]{\mathrm{SU}(#1)_{\mathrm{#2}}}		

\newcommand{\Sl}[3]{(\tilde{L}^{ #1} )^{ #2 }_{ #3 }}
\newcommand{\SlS}[3]{(\tilde{L}^*_{ #1} )_{ #2 }^{ #3 }}

\newcommand{\SQL}[3]{({\tilde{Q}_\mathrm{L}}{}^{ #1} )^{ #2 }_{ #3 }}
\newcommand{\SQR}[3]{({\tilde{Q}_\mathrm{R}}{}^{ #1} )^{ #2 }_{ #3 }}

\newcommand{\LL}[3]{(L^{ #1} )^{ #2 }_{ #3 }}
\newcommand{\QL}[3]{({Q_\mathrm{L}}{}^{ #1} )^{ #2 }_{ #3 }}
\newcommand{\QR}[3]{({Q_\mathrm{R}}{}^{ #1} )^{ #2 }_{ #3 }}

\newcommand{\HH}[3]{(\tilde{H}^{ #1} )^{ #2 }_{ #3 }}
\newcommand{\Ll}[2]{({\tilde{l}_\mathrm{L}}{}^{ #1} )^{ #2 }}
\newcommand{\Lr}[2]{({\tilde{l}_\mathrm{R}}{}^{ #1} )_{ #2 }}

\newcommand{\hh}[2]{\tilde{h}^{ #1}_{ #2 }{}}

\title{Charged scalars from $\SU{3}{}^3$ theories}

\ShortTitle{Charged scalars from $\SU{3}{}^3$ theories}

\author{\speaker{Jos\'e Eliel Camargo-Molina}, Jonas Wess\'en, Roman Pasechnik\\
        \\
        E-mail: \email{Eliel@thep.lu.se}}

\abstract{Inspired by a recently proposed GUT model based on the trinification ($\SU{3}{}^3$) gauge group with a global family ($\SU{3}{F}$) symmetry, we consider an effective low-energy three Higgs doublet model that may shed light on what underlies the observed fermion mass hierarchies and CKM mixing. We discuss possibilities for charged scalars coming from this model to show in collider experiments and show some interesting benchmark points.}

\FullConference{Prospects for Charged Higgs Discovery at Colliders\\
		3-6 October 2016\\
		Uppsala, Sweden}

\begin{document}

\section{Introduction}
In \cite{Camargo-Molina:2016bwm, Camargo-Molina:2016yqm},  a set of Grand Unified Theories (GUTs) based on the trinification $\SU{3}{}^3$ gauge group was proposed. The theories have very promising features that might lead to low energy scenarios that can predict e.g.~the Cabibbo structure for quark mixing and pave the way for an explanation to the mass hierarchies in the SM. The first version \cite{Camargo-Molina:2016yqm}, leads to universality of Yukawa and gauge couplings for chiral quarks and leptons in a supersymmetric (SUSY) scenario. The sectors of light Higgs bosons and leptons are unified into a single chiral super-multiplet dramatically reducing the parameter space of the model while leading to realistic low energy scenarios. In \cite{Camargo-Molina:2016bwm}, it was shown that for a non-SUSY GUT with a very low number of free parameters, it is still possible to generate the theory's different scales by means of radiative symmetry breaking. One of the most interesting low energy scenarios of the theories is a three Higgs doublet model (3HDM). In this work we will understand how a specific low energy 3HDM scenario arises from the model proposed in \cite{Camargo-Molina:2016bwm} and whether charged scalars from the model would be visible in the most common searches at collider experiments. 
\section{From trinification to 3HDM}
Here we will briefly introduce the non-SUSY trinification scenario discussed in \cite{Camargo-Molina:2016bwm}. The model consists of three scalar ($ \tilde{L}, \tilde{Q}_\mathrm{L}, \tilde{Q}_\mathrm{R} $) and three fermion multiplets ($ L, Q_\mathrm{L}, Q_\mathrm{R} $) of the trinification gauge group (and the associated gauge bosons) enhanced with a $\SU{3}{F}$ global symmetry. The final symmetry group of the theory is thus:  
\begin{equation}\label{eq:trinigaugegroup}
G_{\mathrm{T}} = \left[ \SU{3}{L} \times \SU{3}{R} \times \SU{3}{C} \right] \ltimes \mathbb{Z}_3 \times \{\SU{3}{F}  \times \U{A} \times \U{B} \} , 
\end{equation}
where the brackets indicate global (including accidental) symmetries of the theory and the discrete $ \mathbb{Z}_3$ leads to gauge coupling unification. Scalar and fermion multiplets share the same quantum numbers as given in Table. \ref{tab:HSmodelT}. The global minimum of the theory is \footnote{In \cite{Camargo-Molina:2016bwm} it was showed that this is the global minimum whenever it exists as a local minimun}
\begin{equation} \label{eq:vevStructure}
\langle \Sl{i}{l}{r} \rangle = \delta^i_3 \delta^l_3 \delta^3_r \frac{v_T}{\sqrt{2}} , 
\end{equation}
where the $l, r , i$ indices correspond to the  $\SU{3}{L,R,F}$ groups respectively and $\langle \tilde{Q}_\mathrm{L} \rangle = \langle \tilde{Q}_\mathrm{R} \rangle = 0$. This can be shown to spontaneously break $G_{\mathrm{T}}$ as 

\begin{equation}\label{eq:triniBreakingPattern}
\scalebox{0.95}[1]{$G_{\mathrm{T}} \rightarrow G_{\mathrm{LR}} \equiv \SU{3}{C} \times \SU{2}{L} \times \SU{2}{R} \times \U{\BL} \times  \{\SU{2}{F} \times \U{X} \times \U{Z} \times \U{B} \}\,$}.
\end{equation}
\begin{table}[H]
  \begin{center}
    \begin{tabular}{ccccc}
     \toprule                     
                        				 &$\SU{3}{L}$ & $\SU{3}{R}$ & $\SU{3}{C}$ & $ \{\SU{3}{F}\}$ \\      
      \midrule       	
       $\tilde{L}, L $         	    & $\bm{3}$				& $\bar{\bm{3}}$	& $\bm{1}$				& $\bm{3}$				 \\
       $\tilde{Q}_\mathrm{L}, \, Q_\mathrm{L}$  	& $\bar{\bm{3}}$		& $\bm{1}$			& $\bm{3}$				& $\bm{3}$ 				 \\
       $\tilde{Q}_\mathrm{R}, \, Q_\mathrm{R}$	& $\bm{1}$				& $\bm{3}$ 			& $\bar{\bm{3}}$		& $\bm{3}$		 \\
      \bottomrule
    \end{tabular}
    \caption{Field content of the GUT-scale trinification model. The fermionic fields are left-handed Weyl fermions.}
    \label{tab:HSmodelT}
  \end{center}
 \end{table}
Depending on the mass spectrum of the theory after SSB, different effective field theories can arise from this breaking. The decomposition of trinification scalar fields in terms of representations of the group $G_{\mathrm{LR}}$ can be written as (note that uppercase indices take values 1 and 2)
\begin{equation}\label{eq:expansionL}
\begin{aligned}
\Sl{i}{l}{r} = &  \delta^i_I \left[ \delta_L^l \delta_r^R \, \HH{I}{L}{R} + \delta_L^l \delta_r^3 \, \Ll{I}{L}+\delta^l_3 \, \delta_r^R \Lr{I}{R} + \delta_3^l \delta_r^3 \, \tilde{\Phi}^I \right] 		\\
&  + \delta_3^i \left[ \delta_L^l \delta_r^R \, \hh{L}{R} + \delta_L^l \delta_r^3 \, \tilde{l}_\mathrm{L}^s {}^L + \delta_3^L \delta_r^R \, \tilde{l}_\mathrm{R}^s{}_R + \delta_3^l \delta_r^3 \, \left( \tilde{\Phi}^s + \frac{v_3}{\sqrt{2}} \right)\right], 
\end{aligned}
\end{equation}
where the gauge Goldstone bosons in $\tilde{L}$ have been removed. Similar decompositions hold for $\SQL{}{}{}, \SQR{}{}{} $ and their fermion counterparts. In \cite{Camargo-Molina:2016bwm}, we considered a low energy scenario where only $\tilde{h}^L_R$ and $(\tilde{l}_\mathrm{R}^i)_R$ are light and thus present in the EFT. In this work we will study an alternative case in which, after trinification breaking, the field $(\tilde{H}^I)^L_R$  are kept light instead of $\tilde{h}^L_R$. After trinification breaking, the SSB chain continues by means of a VEV in $\tilde{l}_\mathrm{R}$: 
\begin{equation}
\langle \Lr{I}{R} \rangle = \delta^I_2 \delta_R^2 \frac{w}{\sqrt{2}}	
\end{equation}
As also shown in \cite{Camargo-Molina:2016bwm} this can happen through radiative breaking triggered by RG running. With this VEV, the symmetry group is further broken as
\begin{equation}\label{eq:GLE}
G_{\mathrm{LR}} \rightarrow G_{\mathrm{EW}}=\left[ \SU{3}{C} \times \SU{2}{L} \times \U{Y} \right] \times \left\lbrace \U{Y_2} \times \U{Y_3} \times \U{D} \times \U{B} \right\rbrace, 
\end{equation}
As $\SU{2}{R}\times\SU{2}{F}$ is broken, $(\tilde{H}^{1,2})^L_{1,2}$ are now distinct fields and in particular, they are allowed to have different masses. In the following, we will explore the low energy scenario (in other words, we will focus on a region of the LR symmetric theory parameter space) where after $\tilde{l}_\mathrm{R}$ acquires a VEV, only three scalars will remain light, namely:
\begin{equation}\label{eq:lowEfields}
H_1{}^L \equiv (\tilde{H}^1)^L_1 \equiv \Sl{1}{L}{1} \quad , \quad H_2{}^L \equiv (\tilde{H}^2)^L_1 \equiv \Sl{2}{L}{1} \quad , \quad H_3{}^L \equiv \epsilon^{L L'} (\tilde{H}^{*}_1)^2_L  \equiv \epsilon^{L L'} \SlS{1}{L'}{2} 
\end{equation}
\begin{table}[H]
  \begin{center}
    \begin{tabular}[t]{cc}
     \toprule                     
                        				 & $(Y,Y_2,Y_3,D,B)$  \\      
      \hline
       $L_1$						&$(-\frac{1}{2}, -\frac{3}{2}, -\frac{1}{6}, -1,0)$		 \\
       $L_2$						&$(-\frac{1}{2}, -\frac{7}{2}, -\frac{7}{6}, -1,0)$		 \\
       $L_3$						&$(-\frac{1}{2}, -2, -\frac{2}{3}, 0,0)$		 \\
       $e_{\mathrm{R}1}$	&$(1, 3, \frac{4}{3}, 0,0)$		 \\
       $e_{\mathrm{R}2}$	&$(1, 1, \frac{1}{3}, 0,0)$		 \\
       $e_{\mathrm{R}3}$	&$(1, \frac{5}{2}, \frac{5}{6}, 1,0)$		 \\
     \bottomrule
    \end{tabular}
    \hfill
    \begin{tabular}[t]{cc}
     \toprule                    
                        				 & $(Y,Y_2,Y_3,D,B)$  \\      
      \hline
       $q_{\mathrm{L}1}$	&$(+\frac{1}{6}, +\frac{5}{3}, +\frac{13}{18}, -\frac{1}{3}, +\frac{1}{3})$		\\
       $q_{\mathrm{L}2}$	&$(+\frac{1}{6}, -\frac{1}{3}, -\frac{5}{18}, -\frac{1}{3}, +\frac{1}{3})$		 	\\
       $q_{\mathrm{L}3}$	&$(+\frac{1}{6}, +\frac{7}{6}, +\frac{2}{9}, +\frac{2}{3}, +\frac{1}{3})$			\\
       $u_{\mathrm{R}1}$ 	&$(-\frac{2}{3}, -\frac{1}{6}, +\frac{1}{9}, -\frac{2}{3}, -\frac{1}{3})$		 	\\
       $u_{\mathrm{R}2}$ 	&$(-\frac{2}{3}, -\frac{13}{6}, -\frac{8}{9}, -\frac{2}{3}, -\frac{1}{3})$		 	\\
       $u_{\mathrm{R}3}$ 	&$(-\frac{2}{3}, -\frac{2}{3}, -\frac{7}{18}, +\frac{1}{3}, -\frac{1}{3})$		 	\\
       $d_{\mathrm{R}1}$ 	&$(+\frac{1}{3}, -\frac{7}{6}, -\frac{5}{9}, -\frac{2}{3}, -\frac{1}{3})$		 		\\
       $d_{\mathrm{R}2}$ 	&$(+\frac{1}{3}, +\frac{1}{3}, -\frac{1}{18}, +\frac{1}{3}, -\frac{1}{3})$			\\
       $d_{\mathrm{R}3}$ 	&$(+\frac{2}{3}, -\frac{11}{6}, +\frac{4}{9}, +\frac{4}{3}, -\frac{1}{3})$			\\
     \bottomrule
    \end{tabular}
    \hfill
        \begin{tabular}[t]{cc}
     \toprule                         
                        				  & $(Y,Y_2,Y_3,D,B)$  \\      
      \hline
       $H_1$							& $(\frac{1}{2},-1,-\frac{2}{3},0,0)$ \\
       $H_2$							& $(\frac{1}{2},1,\frac{1}{3},0,0)$ \\
       $H_3$							& $(\frac{1}{2},0,\frac{1}{3},0,0)$ \\
    \bottomrule    
    \end{tabular}
    \caption{Lepton (left), quark (middle) and higgs (right) $\U{}$ charges. }
    \label{tab:U1Charges}
  \end{center}
\end{table}
\section{The Model}
We describe here the three Higgs Doublet Model (3HDM) inspired by flavour enhanced non-SUSY trinification. That is, we take the Standard Model (SM) fermion content and the fields in eq. \eqref{eq:lowEfields}, and give them corresponding quantum numbers under the symmetries of $G_{\mathrm{EW}}$ shown eq. \eqref{eq:GLE}. The low energy scenario presented here corresponds to the largest continuous symmetry possible on top of the SM gauge group, $\U{Y_2} \times \U{Y_3}$  \cite{Keus:2013hya}. By fixing the charges of the scalars to that in \cite{Keus:2013hya} (right-most panel in Table \ref{tab:U1Charges}), we find the charges of the fermion fields (left and middle panel of Table \ref{tab:U1Charges}). These charges can all be written as linear combinations of the generators of $G_{\mathrm{T}}$ in Eq.~\eqref{eq:trinigaugegroup}, which are left unbroken by $v_T$ and $w$.
The scalar potential with all the symmetries accounted for is quite simple, 
\begin{equation}
V = - \mu_i^2 |H_i|^2 + \frac{\lambda_{ij}}{2} |H_i|^2 |H_j|^2 + \frac{\lambda'_{ij}}{2} |H_i^\dagger H_j|^2 \qquad \mbox{with}  \qquad
H_i = \begin{pmatrix}
H^+_i \\
\frac{1}{\sqrt{2}}\left(v_i + H^0_i + \i A^0_i \right).
\end{pmatrix}
\end{equation}
We can take all parameters real and  $\lambda_{ij}=\lambda_{ji}$, $\lambda'_{ij} = \lambda'_{ji}$ and $\lambda'_{11}=\lambda'_{22}=\lambda'_{33}=0$ without loss of generality. Assuming $v_{1,2,3} \neq 0$, the extremal conditions are solved by setting $\mu_i^2 = \frac{1}{2}\left(\lambda_{ij} + \lambda'_{ij} \right) v_j^2$ which leads to 
\begin{equation}
V \ni  \frac{1}{2} H^0_i (M_\mathrm{N}^2)_{ij} H^0_j + H^-_i (M_\mathrm{C}^2)_{ij} H^+_j 
\end{equation}
where $ (M_\mathrm{N}^2)_{ij}  = 2 (\lambda_{ij} + \lambda'_{ij} ) v_i v_j $,  $(M_\mathrm{C}^2)_{ij}  = \lambda'_{ij} v_i v_j - \delta_{ij} \sum_k \lambda'_{ik} v_k^2 $ and $ H^-_i \equiv (H^+_i)^*$. The pseudo-scalars $A^0_i$ all become Goldstone modes and are therefore massless.
The trinification fermion fields correspond to the SM ones in the following way
\begin{equation}
L_i{}^L \equiv \LL{i}{L}{3} \,, \,\, e_{\mathrm{R}i} \equiv \LL{i}{3}{1}\,, \,\, q_{\mathrm{L}i}{}^L \equiv \epsilon^{LL'} \QL{i}{}{L'} \,, \,\,u_{\mathrm{R}i} = \QR{i}{1}{} \,, \,\, d_{\mathrm{R}i} = \left( \QR{2}{2}{}, \QR{3}{2}{}, \QR{3}{3}{} \right)^i, 
\end{equation}
leading to the $\U{}$ charges in Table~\ref{tab:U1Charges}. The following Yukawa interactions are allowed by the low energy theory symmetries:
\begin{eqnarray} \label{eq:Yukawas}
\mathcal{L} &\ni& y_{1}^\mathrm{l} H_1^\dagger L_1 e_{\mathrm{R}3} + y_{2}^\mathrm{l} H_1^\dagger L_3 e_{\mathrm{R}1} + y_{3}^\mathrm{l} H_2^\dagger L_2 e_{\mathrm{R}3}+y_{4}^\mathrm{l} H_2^\dagger L_3 e_{\mathrm{R}2}  \\ \nonumber
& & + y_{1}^\mathrm{u} H_1 q_{\mathrm{L}2} u_{\mathrm{R}3} + y_{2}^\mathrm{u} H_1 q_{\mathrm{L}3} u_{\mathrm{R}2} + y_{3}^\mathrm{u} H_2 q_{\mathrm{L}1} u_{\mathrm{R}3} + y_{4}^\mathrm{u} H_2 q_{\mathrm{L}3} u_{\mathrm{R}1} 		\\ \nonumber
& & + y_{1}^\mathrm{d} H_3^\dagger q_{\mathrm{L}2} d_{\mathrm{R}2} + y_{2}^\mathrm{d} H_3^\dagger q_{\mathrm{L}3} d_{\mathrm{R}1} + \mbox{h.c.}
\end{eqnarray}
Its important to note that although the Yukawa sector looks somewhat complicated in comparison to that of the SM, in eq.\eqref{eq:Yukawas}, $y_{i}^a$ are complex numbers and not $3 \times 3$ matrices. In total, the model has 25 free parameters: $\mu_i^2$, $ \lambda'_{ij} $, $‹\lambda'_{ij}$, $y_{1,2,3,4}^\mathrm{u,l}$, $y_{1,2}^\mathrm{d} $ and $g_{1,2,3}$.
\subsection{Cabibbo structure of CKM-matrix}
In the low-energy 3HDM, the quark mass terms in the Lagrangian are 
\begin{equation}
\mathcal{L} = M_{\mathrm{U}}^{ij} \bar{u}_{\mathrm{R}}^i u_{\mathrm{L}}^j + M_{\mathrm{D}}^{ij} \bar{d}_R^i d_{\mathrm{L}}^j +  \mbox{c.c.} \quad \mbox{where}  \quad M_\mathrm{U} = \frac{1}{\sqrt{2}} \footnotesize \begin{pmatrix}
0 & 0 & v_2 y^\mathrm{u}_4		\\
0 & 0 & v_1 y^\mathrm{u}_2 		\\
v_2 y^\mathrm{u}_3  & v_1 y^\mathrm{u}_1 & 0
\end{pmatrix}, \quad	
M_\mathrm{D} = \frac{1}{\sqrt{2}}  \footnotesize \begin{pmatrix}
0 & 0 & v_3 y_2^\mathrm{d} 			\\
0 & v_3 y_1^\mathrm{d} & 0 			\\
0 & 0 & 0
\end{pmatrix},
\end{equation}
giving a massless first generation. We'll arrange the mass eigenstates as
\begin{equation}
 \tilde{u}_\mathrm{R,L}^i = \left(
 u_\mathrm{R,L}, 
 c_\mathrm{R,L},
 t_\mathrm{R,L}		
\right)^i	 , \quad \tilde{d}_\mathrm{R,L}^i  = \left(
 d_\mathrm{R,L},
 s_\mathrm{R,L},
 b_\mathrm{R,L}
\right)^i .
\end{equation}
These are related to the charge eigenstates by unitary transformations,
\begin{equation}
u_\mathrm{L}^i = (V_\mathrm{L}^\mathrm{u} )^{ij} \tilde{u}_\mathrm{L}^i \quad , \quad u_\mathrm{R}^i = (V_\mathrm{R}^\mathrm{u} )^{ij}  \tilde{u}_\mathrm{R}^j \quad , \quad d_\mathrm{L}^i = (V_\mathrm{L}^{\mathrm{d}})^{ij} \tilde{d}_\mathrm{L}^j \quad , \quad d_\mathrm{R}^i = (V_\mathrm{R}^\mathrm{d} )^{ij} \tilde{d}_\mathrm{R}^j .
\end{equation}

From the structure of the Yukawa sector of the model, we find that $V_\mathrm{L}^\mathrm{u,d}$ are such that the CKM matrix automatically takes the Cabibbo form by taking $\tan \theta_\mathrm{C} = \frac{v_2 y_3^\mathrm{u}}{v_1 y_1^\mathrm{u}}$, namely, 
\begin{equation}
V_\mathrm{CKM} = V_\mathrm{L}^\mathrm{u}{}^\dagger V_\mathrm{L}^\mathrm{d} = \footnotesize \begin{pmatrix}
 c_{\theta_\mathrm{C}} & s_{\theta_\mathrm{C}} & 0 \\
 - s_{\theta_\mathrm{C}} & c_{\theta_\mathrm{C}} & 0 \\
 0 & 0 & 1 
\end{pmatrix}.
\end{equation}
\section{Charged scalars at colliders}
The purpose of this work is to explore whether charged scalars coming from the above described 3HDM could be visible at current collider experiments in the most common channels and to trigger the discussion for possible ways of testing the model with available data. The study of neutral scalars and other possible signals is left for further work.
In this light, we calculated branching ratios for the channels usually looked at the LHC in charged scalar searches for five benchmark points (shown in table \ref{tab:BR}). The benchmark points were found through a grid scan of the model parameter space requiring a Higgs-like boson with a mas of $125 \pm 5$ GeV, a stable vacuum at tree level and the correct gauge boson masses in agreement with experiment\footnote{The grid scan found 5 points in around two weeks, which have motivated us to search for more efficient methods to scan the parameter space. Future work will describe one such method based on a genetic algorithm.}. In addition we calculated, for the same benchmark points, the inclusive hadronic cross sections shown in tables \ref{tab:XSHW} and \ref{tab:XSHU} for associated production of charged scalars with $W$ bosons and up-type quarks respectively.  
\subsection{Branching ratios}
\begin{table}[H]
  \begin{center}
   \resizebox{\textwidth}{!}{%
   \footnotesize
    \begin{tabular}{cccccc}
     \toprule                     
                        																	 &$m_{H_1^+}=162$ &$m_{H_1^+}=165$ & $m_{H_1^+}=164$& $m_{H_1^+}=118$&$m_{H_1^+}=138$ \\      
                        																          &$m_{H_2^+}=120$ &$m_{H_2^+}=100$ & $m_{H_2^+}=95$& $m_{H_2^+}=88$&$m_{H_2^+}=135$ \\
      \midrule       	
       BR($\, t \rightarrow H^+ b$)         	    										           &$1.972 \times 10^{-11}$ 	& $5.176 \times 10^{-9}$&$8.338 \times 10^{-9}$  & $3.505\times 10^{-7}$  &   $1.091\times 10^{-8}$         	 \\ 
                                                                       											  & $5.310 \times 10^{-8}$ 	&$4.892 \times 10^{-8}$ &  $3.761 \times 10^{-7}$& $2.536\times 10^{-7}$  &   $3.106\times 10^{-7}$               \\						  
            \midrule                                                                       																									%
       BR($\, H^+ \rightarrow c \bar{s}$)    	                                                                                   	           & $0.997$				& $0.999$&$0.991$&   $0.999$ & $0.973$ \\
           	                                                                                      							   & $0.999$                               & $0.999$&$0.999$&   $0.999$ & $0.780$ \\
              \midrule         	                                                                                      																	    %
       BR($\, H^+ \rightarrow W^+ H^0_i$)    	                                                                               		    &$0.002$				&$0.001$        & $0.008$& Kin. Forb. & $0.026$\\
                                                                     	                                                                                        & Kin. Forb.				&$Kin. Forb.$ & Kin. Forb.&Kin. Forb.   &$0.220$  \\
               \midrule                                                                  	                                                                                     		    %
      \footnotesize BR($\, t \rightarrow H^+ b$)   $\times$  BR($\, H^+ \rightarrow c \bar{s}$)	    	              & $1.094 \times 10^{-11}$    &$5.169 \times 10^{-9}$& $8.264 \times 10^{-9}$&$3.505\times 10^{-7}$    &    $1.061\times 10^{-8}$    \\  
                                                                     	                                                                                          &$5.306\times 10^{-8}$       &$4.891 \times 10^{-8}$& $3.761 \times 10^{-7}$& $2.536\times 10^{-7}$    &   $2.424\times 10^{-7}$   \\
                 \midrule                                                                	                                                                                     		    %
      \footnotesize BR($\, t \rightarrow H^+ b$)   $\times$  BR($\, H^+ \rightarrow c \bar{b}$)		      &$4.336 \times 10^{-18}$    &$1.630 \times 10^{-15}$&$4.007 \times 10^{-15}$& $1.286\times 10^{-13}$    &  $2.200\times 10^{-19}$    \\
                                                                     	                                                                                          &$7.221 \times 10^{-12}$    &$3.670 \times 10^{-14}$&$2.238\times 10^{-14}$& $3.036\times 10^{-15}$  	&   $2.362\times 10^{-15}$   \\
                    \midrule                                                             	                                                                                     		    %
       \footnotesize BR($\, t \rightarrow H^+ b$)   $\times$  BR($\, H^+ \rightarrow \tau^* \nu$)		      &$5.474 \times 10^{-15} $     &$2.517 \times 10^{-13}$ &$2.967 \times 10^{-13}$ &$1.144\times 10^{-11}$  &  $3.212\times 10^{-13}$    \\  
                                                                      	                                                                                          &$2.655 \times 10^{-11}   $   & $2.381 \times 10^{-12}$& $1.350 \times 10^{-11}$& $8.278\times 10^{-12}$&  $7.336\times 10^{-12}$     \\
      \bottomrule
    \end{tabular}}
    \caption{Branching ratios for benchmark points. Masses are in GeV. }
    \label{tab:BR}
  \end{center}
 \end{table}
In table \ref{tab:BR} we show the branching ratios most commonly used in charged scalar searches at the LHC. As an example, both ATLAS \cite{Aad:2014kga} and CMS \cite{Khachatryan:2015qxa} have presented limits for the production of $H^+\rightarrow \tau^* \nu$. The current limits set  BR($\, t \rightarrow H^+ b$)   $\times$  BR($\, H^+ \rightarrow \tau^* \nu$)  to around $10^{-3}$ -- $10^{-2}$ (depending on the charged scalar mass), much above the values calculated for the 3HDM presented in this work. The same holds true for $H^+\rightarrow c \bar{s} , c \bar{b}$  (se e.g.~\cite{ATLAS:2011yia, Khachatryan:2015uua }) and thus for the remaining branching ratios in table \ref{tab:BR} \footnote{The $H^+\rightarrow tb$ channel is kinematically suppressed in all benchmark points}. 
\subsection{Inclusive hadronic cross sections } 
An interesting quantity to look for is the inclusive hadronic cross sections for associated charged scalar production at collider experiments. Being a highly model dependent quantity, it might provide some insight into possible model-exclusive signals and how it compares to other popular models. In tables \ref{tab:XSHU} and \ref{tab:XSHW} we show the cross sections for associated production with up-type quarks and $W$ bosons respectively. Comparing with similar studies performed e.g.~in \cite{Enberg:2011ae} and \cite{Weydert:2009vr}, we find that the signal from charged scalars in this model is several orders of magnitude below MSSM and NMSSM predictions and Standard Model background at the LHC.   
\setlength{\tabcolsep}{3pt}
\begin{table}[H]
  \begin{center}
  \begin{minipage}{0.45\textwidth}
  \resizebox{1.1\textwidth}{!}{%
    \begin{tabular}{cccccc}
     \toprule                     
   $m_{H_1^+}=162$ &$m_{H_1^+}=165$ & $m_{H_1^+}=164$& $m_{H_1^+}=118$&$m_{H_1^+}=138$ \\      
  $m_{H_2^+}=120$ &$m_{H_2^+}=100$ & $m_{H_2^+}=95$& $m_{H_2^+}=88$&$m_{H_2^+}=135$ \\
      \midrule       	
      	 $0.104\times 10^{-10} $   &$0.109 \times 10^{-7}$ 	    & $0.130\times 10^{-7}$    &  $0.367\times 10^{-7}$	&  $0.177\times 10^{-8}$	 \\ 
       	$0.568\times 10^{-8}	  $    & $0.456\times 10^{-8}$    &   $0.352\times 10^{-7}$   &   	$0.240\times 10^{-7} $ &   $0.452\times 10^{-7}$ \\						                  
      \bottomrule
    \end{tabular}}
    \caption{Inclusive hadronic cross sections (in pb) for $H^- / W^+$ associated production (Fig.~\protect\ref{fig:DiagramsXSW})}
    \label{tab:XSHW}
    \end{minipage}
    \qquad
     \begin{minipage}{0.45\textwidth}
     \resizebox{1.1\textwidth}{!}{%
      \begin{tabular}{cccccc}
     \toprule                     
   $m_{H_1^+}=162$ &$m_{H_1^+}=165$ & $m_{H_1^+}=164$& $m_{H_1^+}=118$&$m_{H_1^+}=138$ \\      
  $m_{H_2^+}=120$ &$m_{H_2^+}=100$ & $m_{H_2^+}=95$& $m_{H_2^+}=88$&$m_{H_2^+}=135$ \\
      \midrule       	
       	$0.124\times 10^{-9}	$	                     & $0.138\times 10^{-6}$	    & 	 $0.162\times 10^{-6}$	   & 	 	$0.358\times 10^{-6}	$&	  $0.191\times 10^{-7}$ \\ 
       	$0.561\times 10^{-7}   $                           &  $0.399\times 10^{-7}$   &  $0.294\times 10^{-6}$    &    	$0.190\times 10^{-6} $ &  $0.487\times 10^{-6}$  \\				                                          
      \bottomrule
    \end{tabular}
    }
    \caption{Inclusive hadronic cross sections (in pb) for up-type quark associated  $H^-$ production (Fig.~\protect\ref{fig:DiagramsXSU})}
    \label{tab:XSHU}
    \end{minipage}
  \end{center}
 \end{table}
 \vspace{-40pt}
 \begin{figure}[H]
   \begin{minipage}[b]{0.5\textwidth}
   \centering
  \includegraphics[width=0.4\linewidth]{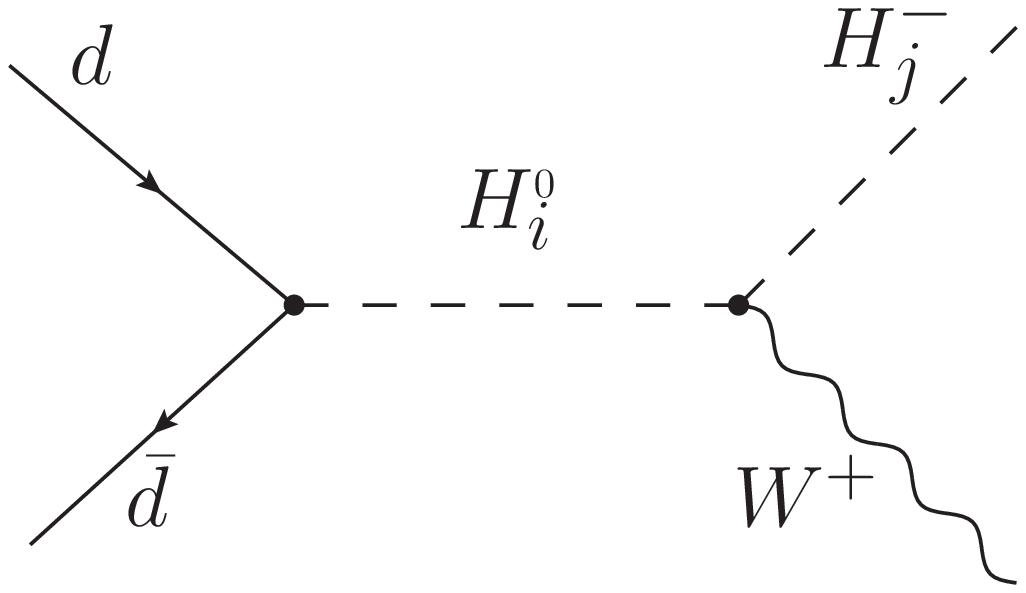}\\ 
  \includegraphics[width=0.35\linewidth]{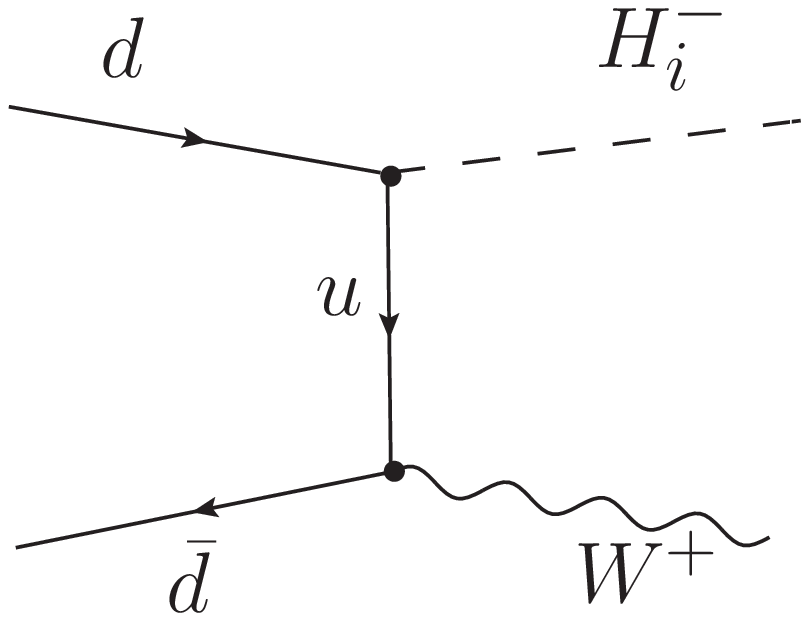}
  \includegraphics[width=0.45\linewidth]{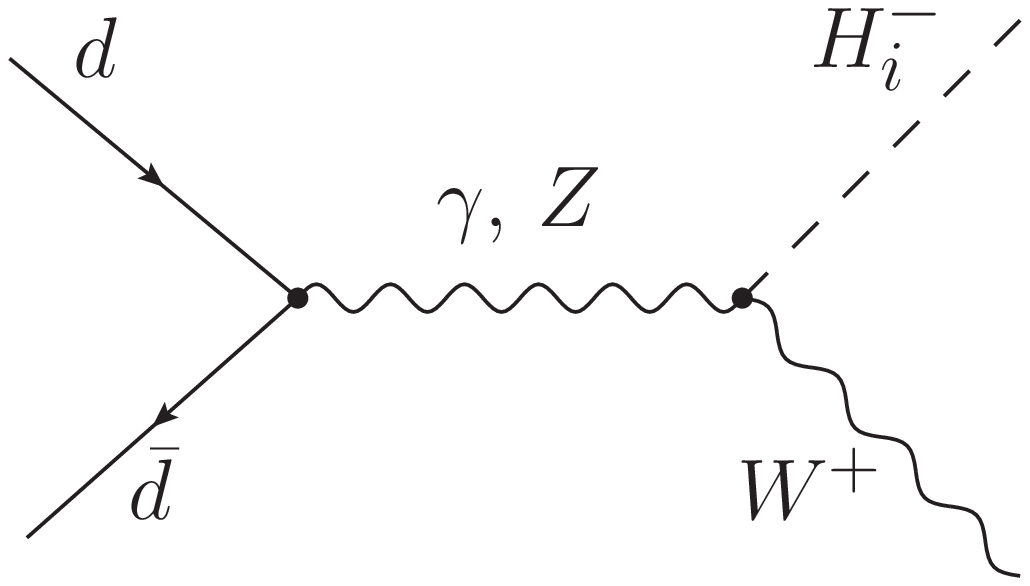} 
  \caption{Diagrams included in the cross section calculation for  $H^- / W^+$ associated production.}
 \label{fig:DiagramsXSW} 
  \end{minipage}
  \hfill
  \begin{minipage}[b]{0.45\textwidth}
  \centering
  \includegraphics[width=0.4\linewidth]{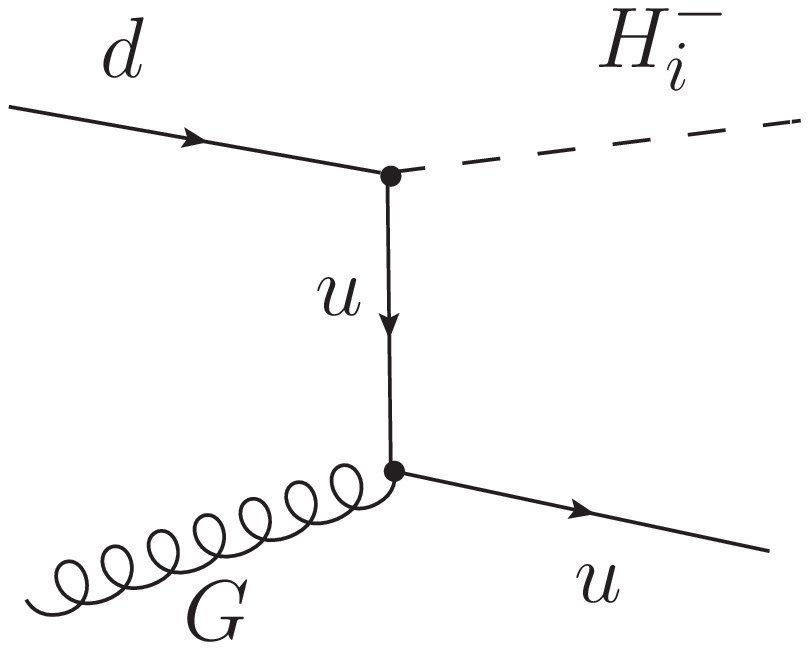}
  \includegraphics[width=0.55\linewidth]{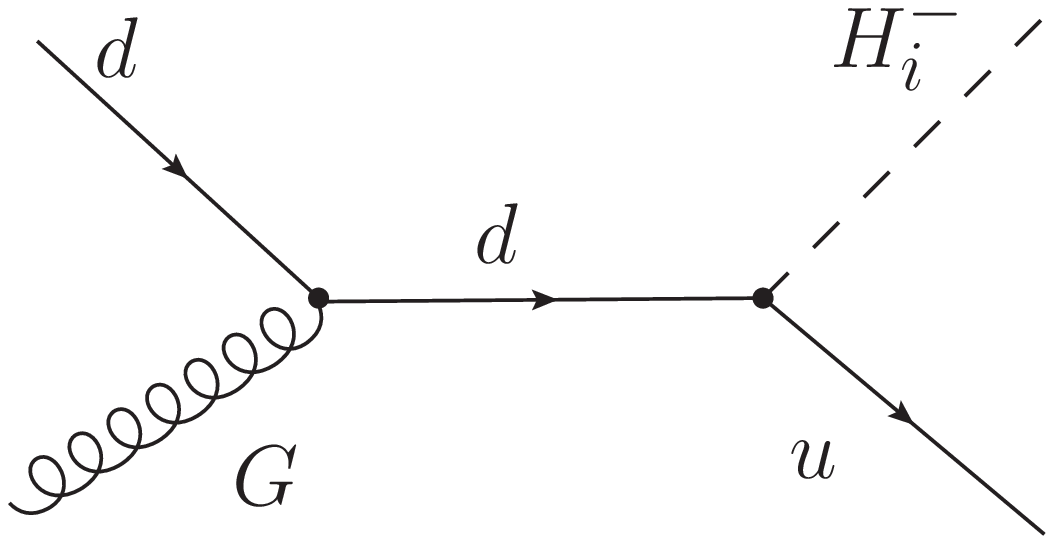} 
  \caption{Diagrams included in the cross section calculation for $H^- / u$ associated production.}
  \label{fig:DiagramsXSU}
    \end{minipage}
\end{figure}
\section{Conclusions and discussion}
By looking at the results, it is clear that the common channels for charged scalar searches will not detect or exclude the charged scalars coming from the proposed model. Due to the unconventional Yukawa sector, a consequence of the high-scale trinification symmetry, the coupling of charged scalars to quarks (controlling the dominant decay channels in e.g.~the Standard Model) leads to very low cross sections and atypical dominant decay channels (e.g. $H^+ \rightarrow c\bar{s}$). This calls for a more thorough phenomenological analysis and the determination of interesting channels to look for in current and future collider experiments.
\footnotesize
\bibliography{bib}
\bibliographystyle{JHEP}

\end{document}